\documentclass[fleqn,usenatbib]{mnras}

\usepackage{newtxtext,newtxmath}

\usepackage[T1]{fontenc}

\DeclareRobustCommand{\VAN}[3]{#2}
\let\VANthebibliography\thebibliography
\def\thebibliography{\DeclareRobustCommand{\VAN}[3]{##3}\VANthebibliography}

\usepackage{graphicx}	
\usepackage{amsmath}	
\usepackage{ulem}
\usepackage[flushleft]{threeparttable}
\usepackage{array} 
\usepackage[english]{babel}
\usepackage{blindtext}
\usepackage{subcaption}

\newcommand{\secref}[1]{\S\ref{#1}}
\newcommand{\bbg}{B_{\rm g}}
\newcommand{\bbo}{B_0}

\title[Bulk motions in guide field magnetic reconnection]{Comptonization by Reconnection Plasmoids in Black Hole Coronae III: Dependence on the Guide Field in Pair Plasma}

\author[S. Gupta, N. Sridhar, {\&} L. Sironi (2023)]{
Sanya Gupta,$^{1}$\thanks{E-mail: sg4038@columbia.edu}
Navin Sridhar,$^{2,3}$\thanks{E-mail: navin.sridhar@columbia.edu}
Lorenzo Sironi$^{2,4}$\thanks{E-mail: lsironi@astro.columbia.edu}
\\$^{1}$Barnard College, Columbia University, 3009 Broadway, New York, NY 10027, USA
\\$^{2}$Department of Astronomy and Columbia Astrophysics Laboratory, Columbia University, 550 W 120th St, New York, NY 10027, USA
\\$^{3}$Cahill Center for Astronomy and Astrophysics, California Institute of Technology, Pasadena, CA 91125, USA
\\$^{4}$Center for Computational Astrophysics, Flatiron Institute, 162 5th Avenue, New York, NY 10010, USA}

\date{Accepted XXX. Received YYY; in original form ZZZ}

\pubyear{2023}

\begin{document}
\label{firstpage}
\pagerange{\pageref{firstpage}--\pageref{lastpage}}
\maketitle

\begin{abstract}
We perform two-dimensional particle-in-cell simulations of magnetic reconnection for various strengths of the guide field (perpendicular to the reversing field), in magnetically-dominated electron-positron plasmas. Magnetic reconnection under such conditions could operate in accretion disk coronae around black holes. There, it has been suggested that the trans-relativistic bulk motions of reconnection plasmoids containing inverse-Compton-cooled electrons could Compton-upscatter soft photons to produce the observed non-thermal hard X-rays. Our simulations are performed for magnetizations $3 \leq \sigma \leq 40$ (defined as the ratio of enthalpy density of the reversing field to plasma enthalpy density) and guide field strengths $0 \leq B_{\rm g}/B_0 \leq 1$ (normalized to the reversing field strength $B_0$). We find that the mean bulk energy of the reconnected plasma depends only weakly on the flow magnetization but strongly on the guide field strength---with $B_{\rm g}/B_0 = 1$ yielding a mean bulk energy twice smaller than $B_{\rm g}/B_0 = 0$. Similarly, the dispersion of bulk motions around the mean---a signature of stochasticity in the plasmoid chain's motions---is weakly dependent on magnetization (for $\sigma \gtrsim 10$) but strongly dependent on the guide field strength---dropping by more than a factor of two from $B_{\rm g}/B_0 = 0$ to $B_{\rm g}/B_0 = 1$. In short, reconnection in strong guide fields ($B_{\rm g}/B_0 \sim 1$) leads to slower and more ordered plasmoid bulk motions than its weak guide field ($B_{\rm g}/B_0 \sim 0$) counterpart.

\end{abstract}

\begin{keywords}
acceleration of particles – black hole physics – magnetic reconnection – relativistic processes – X-rays: binaries
\end{keywords}



\section{Introduction} \label{introduction}
Stellar-mass black holes are observed in ``soft'' and ``hard'' X-ray states. The high-energy, non-thermal X-rays are detected typically during the early-time onset and late-time fading of black hole binary outbursts (i.e., hard states). This emission is traditionally attributed to the unsaturated Comptonization of soft photons by the \textit{corona}, a cloud of hot electrons with typical temperatures of $\sim{\cal O}(100)$\,keV (\citealt{Bisnovatyi-Kogan1977, Dove1997, ZG04}). Observations of X-ray spectral and temporal properties have indicated that the coronal properties (e.g., temperature, geometry, location, size) evolve depending on the phase of the outburst \citep{K19, Sh20, Connors+21, Wang+22}. Yet, the emission mechanism that powers the hard X-rays is still largely unknown.

Magnetic reconnection has been suggested as a mechanism for heating and accelerating electrons in black hole coronae \citep{Galeev1979, DiMatteo1997, B99, Merloni2001a, Merloni2001b, LMS02}, especially in the ``relativistic'' regime where the magnetic energy density is larger than the particle rest-mass energy density \citep{L05}. Recently, \citet{B17} proposed that the trans-relativistic bulk motions of reconnection plasmoids---i.e., magnetic islands / flux ropes resulting self-consistently from the fragmentation of the reconnection layer---could Comptonize the soft disk photons to produce the non-thermal X-ray emission.\footnote{Alternatively, \cite{groselj_23} performed radiative particle-in-cell simulations of turbulence in plasmas of moderate optical depth and showed that most of the turbulence power is transferred directly to the photons via bulk Comptonization,
shaping the peak of the emission around 100 keV. 
} The relative contribution of the particles' internal vs bulk motions to the Comptonized emission can be assessed only with particle-in-cell (PIC) simulations of radiative reconnection, including inverse Compton losses. Yet, most PIC simulations of relativistic reconnection have been conducted in the regime of negligible radiative losses \citep[e.g.,][]{ZH01, LL08, KMS13, G14, G19, SS14, SPG15, S16, werner_16,WU17, PS18, H21, ZSG21, sironi_22,Zhang23}. 


Among the few PIC studies of inverse Compton-cooled reconnection \citep{Werner_2018, SB20, SSB21, SSB23}\footnote{\citet{Chernoglazov23} recently performed PIC simulations of relativistic magnetic reconnection with synchrotron cooling.}, the first three focused on electron-positron plasmas while the latter focused on electron-ion plasmas. \cite{SSB23} confirmed that, regardless of the plasma composition, the bulk motions of the plasmoid chain dominate the inverse Compton power in the regime of strong cooling. Previous papers in this series \citep{SB20,SSB21, SSB23} had studied the properties of plasmoid bulk motions assuming a weak ``guide field'' $B_{\rm g}/B_0 = 0.1$, where $B_{\rm g}$ is the strength of the guide field perpendicular to the reversing field $B_0$.


In this work, we extend the previous papers in this series \citep{SB20, SSB21, SSB23} and study the dependence on the guide field strength, by considering $B_{\rm g}/B_0=0, 0.1, 0.3, 0.6, 1$. This paper focuses on the effect of the guide field on the plasmoid bulk motions, to understand the processes that dominate Comptonization in black hole coronae (note that \citealt{WU17} and \citealt{R19} studied the influence of guide fields on particle heating and non-thermal acceleration).

We focus on the relativistic regime where the magnetization $\sigma$ is larger than unity, which is likely representative of plasma conditions in black hole coronae \citep{B17}. We parameterize the reversing field $B_0$ by the magnetization $\sigma$, which we define as the ratio of magnetic enthalpy density to plasma enthalpy density,
\begin{equation}
 \sigma = \frac{B_0^2}{4\pi n_0 m_{\rm e} c^2} = \left(\frac{\omega_{\rm c}}{\omega_{\rm p}}\right)^2,
\end{equation}
where $n_0$ is the particle density, $\omega_{\rm c} = eB_0/m_{\rm e}c$ is the Larmor frequency, and $\omega_{\rm p} = \sqrt{4\pi n_0e^2/m_{\rm e}}$ is the plasma frequency. We choose not to include the guide field in our definition of $\sigma$, which then quantifies the energy per particle available for dissipation (guide fields get just compressed, and do not transfer energy to the particles).

This paper is organized as follows. In Section \ref{simulation}, we describe the simulation setup. In Section \ref{results}, we present our results, focusing on reconnection bulk motions. Finally, our conclusions, the implications of our work, and future steps are outlined in Section \ref{conclusions}.

\section{PIC Simulation Setup} \label{simulation}

Our simulations are performed with the 3D particle-in-cell code \texttt{TRISTAN-MP} \citep{S05} and we use a Vay pusher \citep{V08} to advance the particle momenta. The setup of the simulations mirrors previous papers in this series \citep{SB20, SSB21, SSB23}---we use a 2D $x-y$ domain, but we track all components of the particles' velocity and of the electromagnetic fields. The reconnection layer is configured by initiating the magnetic field in a ``Harris equilibrium'', ${\bmath B_{\rm in}} = B_0\hat{x}\tanh(2\pi y/\Delta)$, where the direction of the in-plane magnetic field reverses at $y = 0$ over a thickness $\Delta = 100\,c/\omega_{\rm p}$. 

In this paper, we consider a range of guide fields of magnitude $B_{\rm g}/B_0=0, 0.1, 0.3, 0.6, 1$, and we also vary the magnetization $\sigma= 3,10,$ and $40$. The corresponding Alfv\'{e}n speeds for each $\sigma$ are defined as $v_{\rm A}/c =\sqrt{\sigma/(1+\sigma)}= {0.87, 0.95, 0.99}$.\footnote{If we were to include the inertia of the guide field in the definition of the Alfv\'{e}n velocity, we would have \citep{Melzani2014}: 
\begin{equation} \label{proper_v}
 v_{\rm A}'= \frac{B_0}{\sqrt{4\pi \rho c^2 +B_0^2+B_{\rm g}^2}}
\end{equation}} We choose these three values of magnetization to ensure some consistency with the previous papers in this series: $\sigma = 10$ was the baseline in \citet{SB20}, $\sigma = 40$ in \citet{SSB21}, and low magnetization cases including $\sigma = 3$ in \citet{SSB23}. We note that the simulations in \cite{SSB23} were conducted for an electron-ion plasma while the simulations in this paper employ an electron-positron plasma. We initialize $n_0 = 4$ particles per cell (including both species), but we have verified that our results are converged with respect to this choice (more in Appendix~\secref{density}). For all of our analyses, we only consider cells with $\geq 4$ particles to ensure sufficient statistics, e.g., when computing bulk motions.
We refer to the table in Appendix~\secref{sim_table} for the complete set of our input numerical and physical parameters. The same table contains some of the results we obtain.

We resolve the electron inertial length / skin depth ($c/\omega_{\rm p}$) with 5 cells. The size of our reference box is $L_{\rm x}/(c/\omega_{\rm p}) = 1680$, where $L_{\rm x}$ is the half-length of the box along the $x$-direction of reconnection outflows (more in Appendices \secref{box}). We evolve our simulations until $t_{\rm sim} \sim 4.2 L_{\rm x}/v_{\rm A}$, or $185,000$ timesteps, for all cases (the numerical speed of light is 0.45 cells/timestep). We use open boundaries for fields and particles along the $x$-direction. The box grows in the $y$-direction as the simulation progresses, allowing for more plasma and magnetic flux to enter the domain. At the end of the simulations, the length of our box along the $y$-axis is similar or slightly larger than $L_{\rm x}$. We also performed smaller simulations with $L_{\rm x}/(c/\omega_{\rm p}) = 840$ for $\sigma = 10$ to confirm convergence with respect to the domain size (more in Appendix~\secref{box}). As discussed in greater detail in Appendix~\secref{boundary}, we find that in strong guide field cases some plasma tends to accumulate near the $x$-boundaries. To overcome this spurious effect, all the analyses in this paper exclude the simulation cells in the vicinity of the $x$-boundaries (more precisely, within a distance of $0.08\,L_{\rm x}$ from each boundary). While the spurious accumulation is significantl only for strong guide fields, we apply this cut to all our simulations for consistency.

In the initial setup of our simulations, the magnetic pressure outside the layer is balanced by particle pressure in the layer. We initiate reconnection by artificially cooling the hot particles near the center of the domain [$(x,y) = (0,0)$] at the initial time. This generates two reconnection fronts, which after $\sim1.5$ Alfv\'{e}n crossing times reach the $x$-boundaries of the computational domain. After this time, reconnection attains a ``quasi-steady state'' (more in the Appendix~\secref{steady_state}). 

\section{Results} \label{results}

\subsection{Structure of the reconnection layer} \label{rec_layer}

\begin{figure*}
 \includegraphics[width=\textwidth]{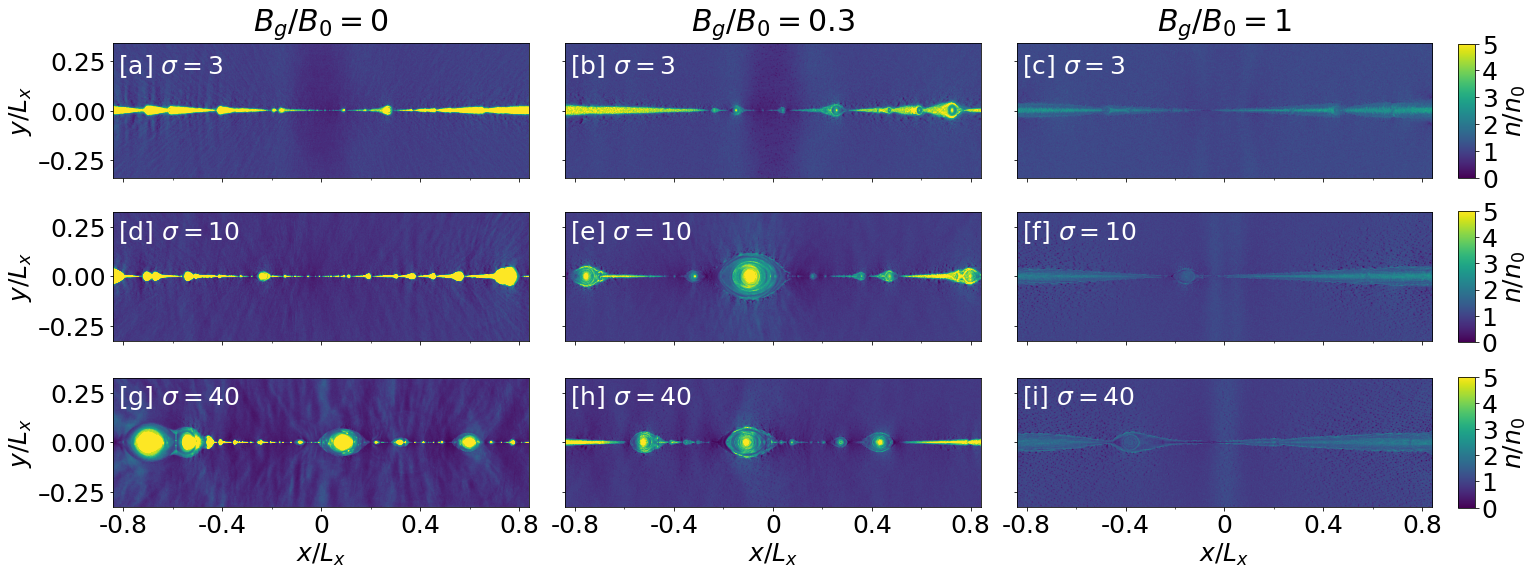}
 \caption{2D snapshots of the reconnection layer at time $Tv_{\rm A}/L_{\rm x} \sim 4$ for magnetizations $\sigma = 3, 10, 40$ (increasing from top to bottom) and guide field strengths $B_{\rm g}/B_0 = 0, 0.3, 1$ (increasing from left to right). All simulations are performed with our fiducial box size $L_{\rm x}/(c/\omega_{\rm p}) = 1680$. The panels show particle number density, $n$, in units of the upstream number density, $n_0$.}
 \label{fig:2d_img}
\end{figure*}

Fig.~\ref{fig:2d_img} shows a snapshot of the reconnection layer for different strengths of guide field and magnetization.\footnote{As described above, the images in Fig.~\ref{fig:2d_img} exclude the simulation cells in the vicinity of the $x$-boundaries (within a distance of $0.08\,L_{\rm x}$ from each boundary). 2D images of the full $x$-extent of the layer are shown in Fig.~\ref{fig:og2d}.} We discuss first the dependence on guide field strength, and then on magnetization.

For a fixed magnetization (e.g., see $\sigma=40$, bottom row in Fig.~\ref{fig:2d_img}, as a representative case), the reconnected plasma is far less compressed in cases with a stronger guide field, since the pressure of the guide field resists compression. Also, the layer is generally thicker for stronger guide fields, consistent with the discussion in 
\citet{Zenitani2008} on the role that guide fields play in regulating the width of the reconnection layer. This has an important consequence: thinner layers---realized for smaller guide fields---are more prone to fragmentation into plasmoids.
This is apparent when comparing the leftmost and rightmost panels in Fig.~\ref{fig:2d_img}. Stronger guide field cases exhibit smoother outflows without many plasmoids (right column). In contrast, weaker guide field cases display a hierarchical chain of plasmoids of various sizes (left column). Smaller plasmoids merge with 
each other and form larger plasmoids (occasionally even \textit{monster plasmoids}\footnote{We define \textit{monster plasmoids} as large plasmoids whose extent is 10-20\% of the total length of the reconnection layer–––similar to the definition in \citet{L12}.}). This is particularly apparent in the left and middle columns: e.g., in panel [g], we see two large plasmoids exiting the simulation box at $x/L_{\rm x} \sim -0.75$ and another large plasmoid near the center at $x/L_{\rm x} \sim 0.1$; in panel [h], there is one large plasmoid near $x/L_{\rm x} \sim -0.1$; in contrast, in panel [i] there are no large plasmoids. Looking specifically at panels [g] and [h], one concludes that the structure of the plasmoid chain for $\bbg/\bbo\lesssim 0.3$ is similar to the case $\bbg/\bbo=0.1$ explored in previous papers in this series \citep{SB20,SSB21,SSB23}, i.e., the layer exhibits a prominent fragmentation into plasmoids.\footnote{As the plasma in our simulations is not radiatively cooled, particles are nearly symmetrically distributed in plasmoids. In contrast, the strongly cooled simulations by \citet{SSB21} showed a non-uniform plasma density distribution inside moving plasmoids, with near-vacuum regions at the front.} 

At low guide field strengths, the dependence on magnetization is 
consistent with previous works in this series: higher $\sigma$ leads to more fragmentation. This is most evident by comparing panels [a], [d], and [g] in the leftmost column (for $B_{\rm g}/B_0=0$). In panel [a], we see small, elongated plasmoids streaming steadily from the central region towards the boundaries, while panel [g] shows large round plasmoids throughout the reconnection layer, merging with each other and moving in a more stochastic way. 

\subsection{Reconnection rate} \label{rec_rate}

\begin{figure*}
	\includegraphics[width=\textwidth]{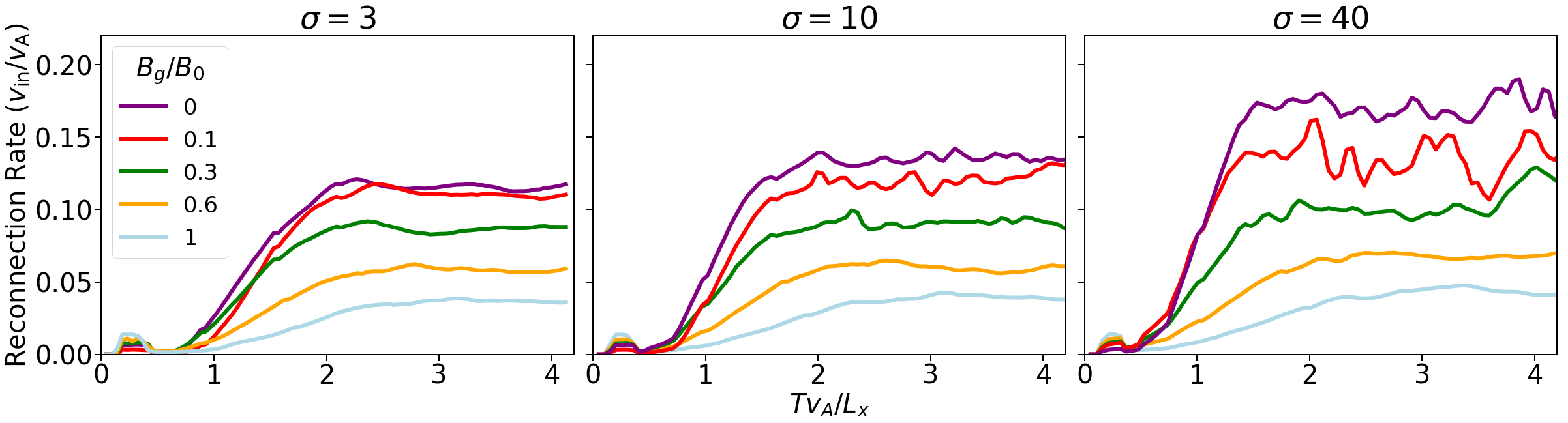}
 \caption{Reconnection rate in units of the Alfv\'{e}n speed, $v_{\rm in}/v_{\rm A}$, as a function of time (in units of $L_{\rm x}/v_{\rm A}$). Colors represent guide field strengths ($B_{\rm g}/B_0$): purple = 0, red = 0.1, green = 0.3, yellow = 0.6, blue = 1. The magnetization increases from left to right panel ($\sigma = 3, \sigma = 10, \sigma = 40$).}
 \label{fig:rec_rate}
\end{figure*} 

 We define the reconnection rate as the upstream plasma's inflow velocity $v_{\rm in}$ into the layer. This rate is computed by taking the spatial average of the inflow velocity, $v_{\rm y}$, over a rectangular box located at $-0.90 \leq x/L_{\rm x} \leq 0.90$ and $0.15 \leq y/L_{\rm x} \leq 0.20$.

For all magnetizations and guide field strengths, we notice a similar time evolution in Fig.~\ref{fig:rec_rate}: a small bump in the reconnection rate at $Tv_{\rm A}/L_{\rm x} \sim 0.2$ (a consequence of our choice for initiating reconnection), followed by an increase until $Tv_{\rm A}/L_{\rm x} \sim 2$, and, finally, a quasi-steady reconnection rate with some fluctuations. The steep increase in the reconnection rate below $Tv_{\rm A}/L_{\rm x} \sim 2$ occurs while the reconnection fronts are on their way from the center to the boundaries (only the plasma between the two fronts is inflowing into the layer). The fluctuations seen during the quasi-steady state (e.g., purple and red lines in the right panel of Fig.~\ref{fig:rec_rate}) are caused by the mergers of large plasmoids and their escape from the domain (see Fig.~\ref{fig:2d_img}). While all cases exhibit some level of fluctuations during the quasi-steady state, we find that the fluctuation amplitude is largest for high $\sigma$ and/or low $B_{\rm g}$. As discussed in the previous section, this is because the tendency for fragmentation into plasmoids is most pronounced for high $\sigma$ and low $B_{\rm g}$.

As seen in Fig.~\ref{fig:rec_rate}, the reconnection rate exhibits a strong dependence on the guide field strength for all magnetizations: cases with strong guide fields have lower reconnection rates than those with weak guide fields. As an example, in the rightmost panel of Fig.~\ref{fig:rec_rate}, we find that the peak reconnection rate (attained momentarily during the quasi-steady state) for $B_{\rm g}/B_0 = 1$ is $v_{\rm in}/v_{\rm A} \sim 0.04$, whereas for $B_{\rm g}/B_0 = 0$ it is much larger, $v_{\rm in}/v_{\rm A} \sim 0.17$. This trend is consistent across all magnetizations, as the $B_{\rm g}/B_0 = 0$ case (purple curve) consistently reaches quasi-steady values that are $3-4\times$ higher than than the $B_{\rm g}/B_0 = 1$ case of the same magnetization (blue curve). This trend also persists if the Alfv\'en speed is defined as in Eq.~\ref{proper_v}. In Table \ref{tab:parameters}, we quote the average reconnection rate during the quasi-steady state for all the simulations of this work.

\subsection{Bulk motion profile} \label{bulk_motion}
In this subsection, we discuss the effect of magnetization and guide field on the bulk motions of the reconnected plasma. For the rest of the paper, 
we define the reconnected plasma as the region where particles starting from above and below the mid-plane ($y = 0$) contribute at least $1\%$ to the mixture \citep{R19}.

\begin{figure*}
 \includegraphics[width=\textwidth]{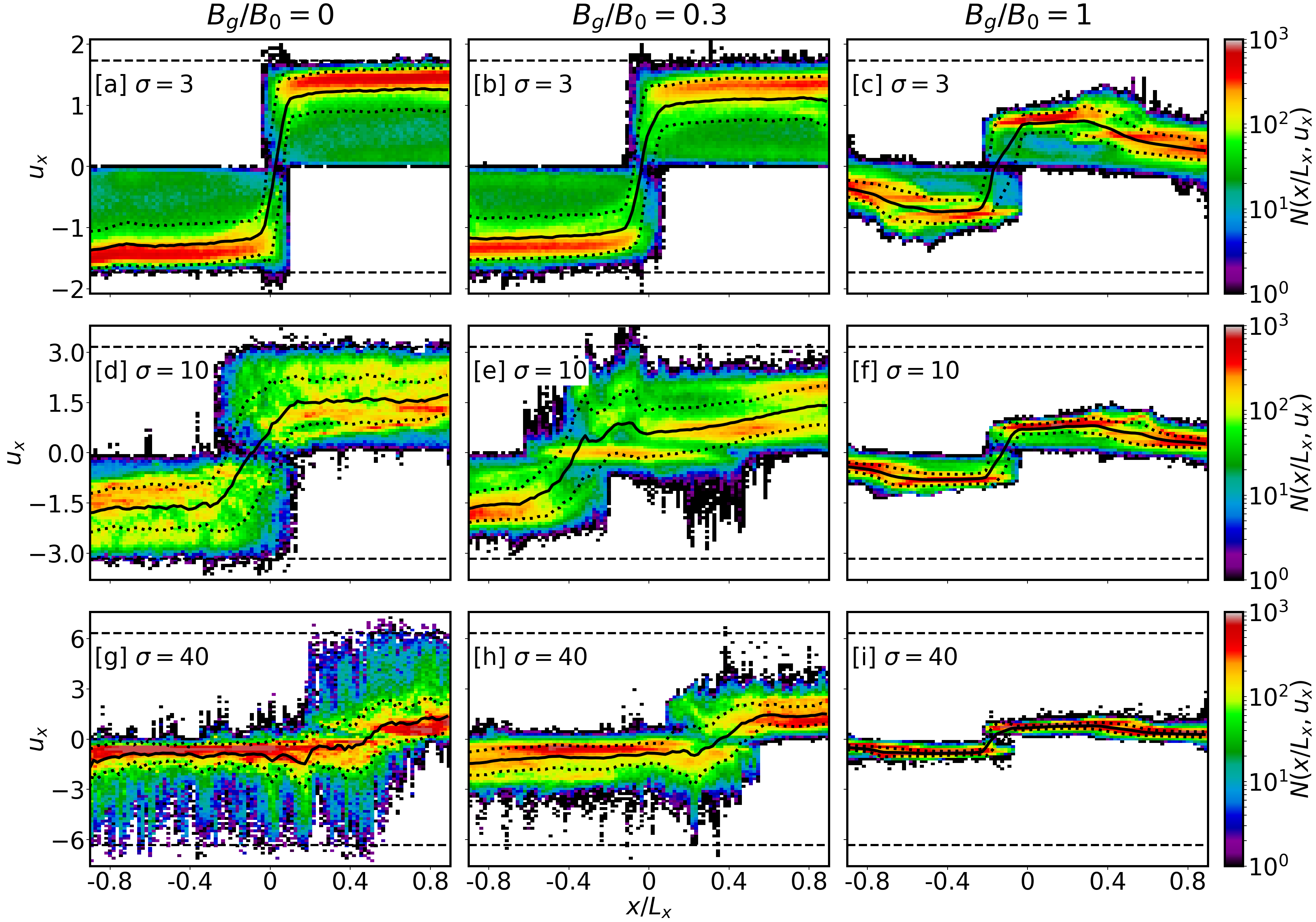}
 \caption{Bulk motions of the reconnected plasma, viewed in the $x-u_{\rm x}$ phase space. Color represents the particle number density. The magnetization increases from top to bottom ($\sigma = 3, \sigma = 10, \sigma = 40$) and the guide field increases from left to right ($B_{\rm g}/B_0 = 0, B_{\rm g}/B_0 = 0.3, B_{\rm g}/B_0 = 1$). The solid black curve in each plot shows the density-weighted mean of $u_{\rm x}$ along the $x$-axis, while the dotted curves show the corresponding standard deviation. The dashed horizontal lines show the Alfv\'{e}n limit, $u_{\rm x}=\pm \sqrt{\sigma}$. All phase space plots are time-averaged over $2 \leq Tv_{\rm A}/L_{\rm x} \leq 4.2$ when the layer is in a quasi-steady state.}
 \label{fig:yux}
\end{figure*}

Bulk motions are calculated as follows. For every cell, the mean particle velocity, $\boldsymbol{\beta}$, is computed as an average over all electrons and positrons in the local patch of neighboring $5 \times 5$ cells \citep{R19}. We then calculate the bulk 4-velocities $u_{\rm x} = \Gamma\beta_{\rm x}$ and $u_{\rm y} = \Gamma\beta_{\rm y}$, where $\Gamma = 1/\sqrt{1 - {\beta}^2}$. 
Here, $u_{\rm x}$ is the component along the reconnection outflow, whereas $u_{\rm y}$ is along the inflow.
The phase-space plots $x-u_{\rm x}$, averaged over the quasi-steady-state $2 \leq Tv_{\rm A}/L_{\rm x} \leq 4.2$, are presented in Fig.~\ref{fig:yux} for different magnetizations and guide fields. 
The array of guide fields and magnetizations presented here mirrors that in Fig.~\ref{fig:2d_img}. The solid black curve in each plot shows the density-weighted mean $\langle u_{\rm x} \rangle$ computed at each $x$, while the dotted curves show the corresponding standard deviation. The dashed horizontal lines show the Alfv\'{e}nic limits, $u_{\rm x}=\pm \sqrt{\sigma}$. For completeness, we also compute the $u_{\rm x}-u_{\rm y}$ phase-space plots, which can be found in Appendix~\secref{bulk_cont}.

Regardless of the strength of the guide field or the magnetization, the general spatial trend of the density-weighted mean $\langle u_{\rm x} \rangle$ is similar: a fast increase away from the central region which then levels off at a nearly constant ``saturation'' value. We now discuss the dependence on magnetization, and then on guide field strength.

At fixed guide field, we notice that magnetization plays a key role in the fraction of reconnected plasma that reaches the Alfv\'{e}nic limit (this also holds true if the Alfv\'{e}n speed is defined as in Eq.~\ref{proper_v}). For higher $\sigma$ cases, the fraction of plasma reaching the Alfv\'{e}nic limit decreases. This effect can be clearly seen by comparing the $\sigma=3$ (panel [b]) and $\sigma=40$ (panel [h]) cases in the middle column of Fig.~\ref{fig:yux}. At fixed guide field, the saturation value of $\langle u_{\rm x} \rangle$ is not strongly dependent on magnetization. However, we notice that different $\sigma$ yield a rather different dispersion of bulk motions around the mean (at a given $x$), suggesting that the layer's stochasticity is strongly dependent on magnetization.
For instance, when comparing panel [g] to panel [a], we see that, at fixed $x$, the reconnected plasma spans a much wider range of $u_{\rm x}$ at higher $\sigma$. This can be quantified by computing the ratio 
$\Sigma_{u_{\rm x}}/\langle u_{\rm x}\rangle$ between the average standard deviation and the average mean (averaged over the region where $\langle u_{\rm x}\rangle$ attains a quasi-constant value).
For panels [a] and [g], we find $\Sigma_{u_{\rm x}}/\langle u_{\rm x}\rangle \sim 0.25$ and $\Sigma_{u_{\rm x}}/\langle u_{\rm x}\rangle \sim 1$, respectively. In short, at higher magnetizations the bulk motions are less likely to reach the Alfv\'{e}nic limit, but they exhibit a wider range of variations (i.e., higher stochasticity).

The guide field strength has a strong influence on the profile of $\langle u_{\rm x}\rangle$. We find that bulk motions are generally slower when increasing the guide field strength, with a negligible fraction of plasma that reaches the Alfv\'{e}nic limit for strong guide fields.
As the guide field increases, bulk motions slow down to trans-relativistic speeds. For instance, in panel [a] ($B_{\rm g}/B_0=0$), the saturation speed is $|\langle u_{\rm x}\rangle| = 1.5$, while in panel [c] ($B_{\rm g}/B_0=1$), the saturation speed is $|\langle u_{\rm x}\rangle| = 1$. In fact, the inertia of the larger guide field leads to slower bulk motions. This is also reflected in the gradient of $\langle u_{\rm x}\rangle$ near the center: $B_{\rm g}/B_0 = 0$ cases reach their saturation speeds closer to the central region than for larger guide fields. The magnetic tension of the reconnected field has a harder time accelerating the plasma in stronger guide field cases, due to the additional inertia of the guide field. 


We see an important trend in the stochasticity of bulk motions when varying the guide field strength. This is most apparent when comparing either panels [a] and [c] or panels [g] and [i]. We see that the dotted curves (denoting standard deviation) in panels [a] and [g], $B_{\rm g}/B_0 = 0$, are much farther from the solid curve (denoting the mean) than in panels [c] and [i], $B_{\rm g}/B_0 = 1$. As the guide field strength increases, the outflow becomes more ordered and the stochasticity of bulk motions significantly drops. As discussed above, this is ultimately related to the fact that layers with stronger guide fields are far less prone to fragmentation into plasmoids.



We also notice that a small fraction of reconnected plasma flows opposite to the mean motion (i.e., we see spikes with $u_{\rm x} < 0$ in places where $\langle u_{\rm x}\rangle > 0$ and viceversa). We see this feature mostly in cases with low guide fields and/or large magnetizations, i.e., where fragmentation into plasmoids is most pronounced. As in previous papers of this series, we interpret the signature of plasma flowing opposite to the mean motion as due to the accretion of a smaller, leading plasmoid by a larger, trailing one. In this case, the large plasmoid pulls back the small plasmoid, which then moves against the mean motion.
For instance, we can map the central panel [e] of Fig.~\ref{fig:yux} with the corresponding panel [e] of Fig.~\ref{fig:2d_img}. In the latter, we see a large plasmoid near the center accreting small plasmoids from its two sides. This is reflected in the two spikes at $x/L_{\rm x} \sim \pm 0.25$ in directions opposite to the mean outflow direction at the same $x$.

We conclude this subsection with a cautionary note. For weak guide fields, the mean bulk speeds at $|x|/L_{\rm x} > 0.2$ (i.e., once they attain their saturation values) are nearly constant. In contrast, for strong guide fields (e.g., panel [c]), we observe faster flows at $0.2 < |x|/L_{\rm x} < 0.6$, followed by a decline when approaching the boundary of the box. We attribute this effect to the artificial accumulation of guide fields and particles near the $x$-boundaries described in Section~\ref{simulation}. This appears to slow down the motions near the boundaries (at $\lvert x \rvert/L_{\rm x}>0.7$) for strong guide fields, especially at later times. However, in Appendix~\secref{steady_state}, we show that the overall trends in the properties of bulk motions reported in this paper are extremely robust during the quasi-steady state, so the late-time slow-down near the boundaries for large guide field cases does not appreciably change our conclusions.


\subsection{Bulk energy spectra} \label{energy} 

\begin{figure*}
 \includegraphics[width=\textwidth]{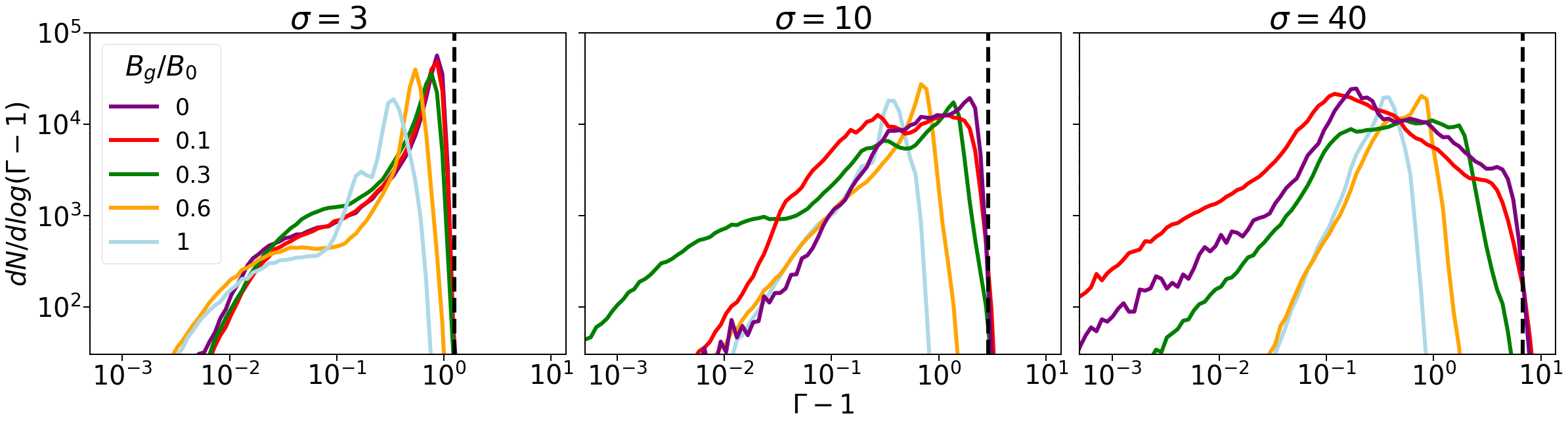}
 \caption{Bulk energy spectra of the reconnected plasma, averaged over $2 \leq Tv_{\rm A}/L_{\rm x} \leq 4.2$. The colors represent guide fields strengths ($B_{\rm g}/B_0$): purple = 0, red = 0.1, green = 0.3, yellow = 0.6, blue = 1. The magnetization increases from left to right ($\sigma = 3, \sigma = 10, \sigma = 40$). The vertical dashed line in each panel shows the Alfv\'{e}nic limit, $\Gamma-1 = \sqrt{1+\sigma}-1$.}
 \label{fig:bulk_spect}
\end{figure*}
We construct particle spectra accounting for bulk energy alone and we present them in 
Fig.~\ref{fig:bulk_spect}. The spectra are time-averaged over the quasi-steady state. 
We test the dependence of the bulk energy spectra on the initial number of computational particles per cell and the size of the simulation box in Appendix~\secref{density} and Appendix~\secref{box}, respectively, to demonstrate convergence of our results. Here, we present the dependence on magnetization and guide field strength.

We find that both the spectral width and the peak location have a strong dependence on magnetization. 
With regard to the former, we see that the bulk spectra get much broader with increasing magnetization, at fixed guide field. E.g., the well-defined, sharp peak in the left panel ($\sigma=3$) for $B_{\rm g}/B_0=0.3$ (green line) transitions to a broad plateau extending from $\Gamma - 1=0.05$ to $\Gamma - 1=5$ in the rightmost panel ($\sigma=40$). This is due to the fact that higher magnetizations have an enhanced tendency for fragmentation into plasmoids, which in turn generates more stochastic motions. With regard to the location of the spectral peak, in the weak guide field cases ($B_{\rm g}/B_0\leq0.1$), the peak of the bulk spectrum is near the Alfv\'{e}nic limit for low magnetizations (left), but it is much lower than the Alfv\'{e}nic limit for high magnetizations (right). In other words, at higher $\sigma$ most of the particles move at bulk speeds well below the Alfv\'{e}nic limit. 

Similar arguments explain the trend with guide field strength, at fixed magnetization. At lower guide fields, the more copious fragmentation into plasmoids 
results in more stochastic motions and in wider bulk spectra; in contrast, the ordered motions we observe for strong guide fields produce sharper-peaked spectra. This holds for $\sigma\gtrsim 10$, e.g., in the rightmost panel ($\sigma=40$), spectra of low guide fields (purple and red curves) are much wider than those for strong guide fields (yellow and cyan). This trend is not observed for our lowest magnetization, $\sigma=3$ (leftmost panel), where motions are rather ordered even for low guide fields, and so the bulk spectrum has a similar shape for all $B_{\rm g}/B_0$. We also notice that the spectral cutoff generally reaches higher energies for weaker guide fields. For instance, at $\sigma=40$ the spectral cutoff for $B_{\rm g}/B_0 = 0$ reaches $\Gamma - 1 \sim \sqrt{\sigma}\sim 6$, whereas for $B_{\rm g}/B_0 = 1$ it is trans-relativistic, $\Gamma-1\sim 1$ (consistent with Fig.~\ref{fig:yux}). In summary, with increasing guide field strength at fixed magnetization, the bulk energy spectra generally shift to lower energies and get narrower.


We separately comment on the low-energy tails ($\Gamma-1 \leq 0.1$) seen in some cases, which we attribute to the presence of larger, slower-moving plasmoids.
For instance, in the middle panel ($\sigma=10$), the low-energy tail of the green spectrum (for $B_{\rm g}/B_0 = 0.3$) is due to the formation of the massive central plasmoid seen in panel [e] of Fig.~\ref{fig:2d_img}.


\begin{figure}
 \includegraphics[width=0.5\textwidth]{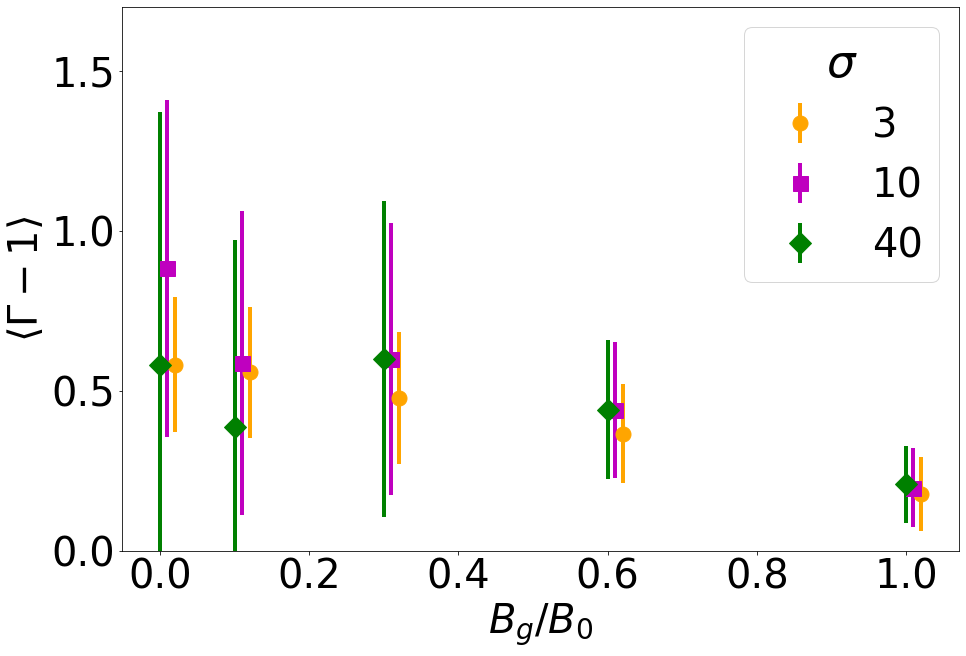} 
 \caption{Bulk motion statistics of the reconnected plasma in models with various $B_{\rm g}/B_0$ and magnetization. Yellow circles, purple squares, and green diamonds indicate mean bulk outflow energy for a range of guide field strengths ($0 \leq B_{\rm g}/B_0 \leq 1$) for $\sigma = 3$, $\sigma = 10$, and $\sigma = 40$, respectively. The error bars indicate the standard deviation. Both mean and standard deviation are time-averaged over the quasi-steady state, $2 \leq Tv_{\rm A}/L_{\rm x} \leq 4.2$.}
 \label{fig:bulkout}
\end{figure}

Fig.~\ref{fig:bulkout} shows the time-averaged and density-weighted $\langle \Gamma-1\rangle$ computed from the spectra in Fig.~\ref{fig:bulk_spect}. The error bars illustrate the dispersion away from the mean bulk motion, as quantified through the standard deviation of the time-averaged spectra in Fig.~\ref{fig:bulk_spect}. We summarize our findings on the dependence of the bulk motions' properties on $\sigma$ and guide field strength here, aided by Fig.~\ref{fig:bulkout}: (1) There is an overall decrease in $\langle{\Gamma -1}\rangle$ with increasing guide field strength, for all magnetizations. $\langle{\Gamma -1}\rangle$ drops by nearly a factor of two from $B_{\rm g}/B_0 = 0$ to $B_{\rm g}/B_0 = 1$. 
(2) While $\sigma = 3$ and 10 have a monotonic decrease in $\langle\Gamma-1\rangle$ with increasing guide field strength, the $\sigma=40$ case shows a decline from $B_{\rm g}/B_0 = 0$ to $B_{\rm g}/B_0 = 0.1$, followed by an increase from $B_{\rm g}/B_0 = 0.1$ to $B_{\rm g}/B_0 = 0.3$, and then a steady decrease for even stronger guide fields. This non-monotonicity is due to the formation of a slow-moving \textit{monster plasmoid} for $B_{\rm g}/B_0 = 0.1$ (akin to, but larger than the plasmoid seen in panel [g] of Fig.~\ref{fig:2d_img}).
(3) The standard deviation $\Sigma_{\Gamma - 1}$, denoted by the error bars, is significantly dependent on both guide field and magnetization. It should be interpreted as a signature of the stochasticity of bulk motions. We find that bulk motions are more stochastic (i.e., less ordered) for smaller $B_{\rm g}/B_0$ and/or larger magnetizations. For $\sigma\gtrsim 10$, the dispersion in bulk motions drops by more than a factor of five from $B_{\rm g}/B_0 = 0$ to $B_{\rm g}/B_0 = 1$.

\section{Conclusions} \label{conclusions}
In this paper, we have investigated with 2D PIC simulations the impact of the guide field strength on relativistic reconnection and, in particular, on the properties of plasmoid bulk motions. Our main results and their implications for astrophysical plasmas are as follows. 

\begin{enumerate}

 \item \textbf{Reconnection rate:} The reconnection rate has a strong dependence on the guide field strength, with the case of zero guide field consistently having the highest reconnection rate. During the quasi-steady state, the reconnection rate in weaker guide fields displays larger temporal fluctuations, which we attribute to a more pronounced fragmentation into plasmoids (see next point).
 
 \item \textbf{Fragmentation:} In stronger guide fields ($B_{\rm g}/B_0 = 1$), the reconnection layer displays little fragmentation, showing smoother, more uniform, and less compressed outflows. In agreement with previous works \citep{WU17,R19}, we find that lower magnetizations result in less fragmented layers. 

 \item \textbf{Bulk energies}: We find that the mean bulk energy depends weakly on the flow magnetization (for $\sigma \gtrsim 3$) and strongly on the guide field strength---with $B_{\rm g}/B_0 = 1$ yielding a mean bulk energy twice smaller than $B_{\rm g}/B_0 = 0$. The dispersion of bulk motions around the mean---a signature of stochasticity in the plasmoid chain---is nearly independent of magnetization for $\sigma \gtrsim 10$, and it is strongly dependent on the guide field strength---dropping by more than a factor of two from $B_{\rm g}/B_0 = 0$ to $B_{\rm g}/B_0 = 0$ (for large magnetizations, $\sigma\gtrsim10$, it drops by more than a factor of five). The bulk energy spectrum has a clear, narrow peak for strong guide field cases, while it is quite broadly-peaked for weak guide field cases.
\end{enumerate}

This paper, as well as previous papers in this series \citep{SB20, SSB21, SSB23}, aims at exploring whether magnetic reconnection can power the observed hard, nonthermal X-rays from the coronae of accreting black holes. 
The Comptonized X-ray emission has a high-energy cutoff at $\sim 100$\,keV. If Comptonization is powered by plasmoid bulk motions, we would require the bulk energy spectrum to extend at least up to $\gtrsim 100$\,keV. We have demonstrated that the mean bulk energy of the reconnected plasma is strongly sensitive to the guide field strength, dropping by roughly a factor of two from $B_{\rm g}/B_0=0$ to $B_{\rm g}/B_0=1$, largely independent of magnetization. It follows that reconnection in strong guide fields might not be able to reproduce the observed $\sim 
 100$~keV peak in the Comptonized X-ray spectrum. We conclude that a scenario based on Comptonization from plasmoid bulk motions in pair plasma 
requires
both a strong magnetization ($\sigma \gtrsim 3$) and a weak guide field strength ($B_{\rm g}/B_0 \lesssim 0.3$), in order to explain the 100\,keV peak seen in X-ray binaries. 

In the future, we plan to extend this work by adding Compton cooling and extracting self-consistent radiative spectra, as a function of magnetization and guide field strength. It will also be useful to investigate how the properties of plasmoid bulk motions change for an electron-ion plasma, as a function of the guide field strength.

\section*{Acknowledgements}

This paper benefited from useful discussions with Luca Comisso and Erin Kara. This project made use of the following computational resources: NASA Pleiades supercomputer as well as the Ginsburg and Terremoto HPC clusters at Columbia University. N.S. acknowledges the support from NASA (grant
number 80NSSC22K0332), NASA FINESST (grant number
80NSSC22K1597), Columbia University Dean’s fellowship,
and a grant from the Simons Foundation. N.S. performed
part of this work at the Aspen Center for Physics, which is
supported by the National Science Foundation grant PHY2210452. This work was supported by a grant from the Simons Foundation (00001470, to L.S.). 
L.S. acknowledges support from DoE Early Career Award DE-SC0023015 and from NSF AST-2108201. This research was facilitated by the Multimessenger Plasma Physics Center (MPPC), NSF grant PHY-2206609 to L.S.
S.G. would like to particularly thank L.S. and N.S. for all their help, support, and encouragement throughout this project. 

\section*{Data Availability}

The data underlying this paper will be shared upon reasonable request to the authors. 



\bibliographystyle{mnras}
\bibliography{gupta22} 




\appendix

\section{Particle Number Density} \label{density}

\begin{figure}
 \includegraphics[width=0.5\textwidth]{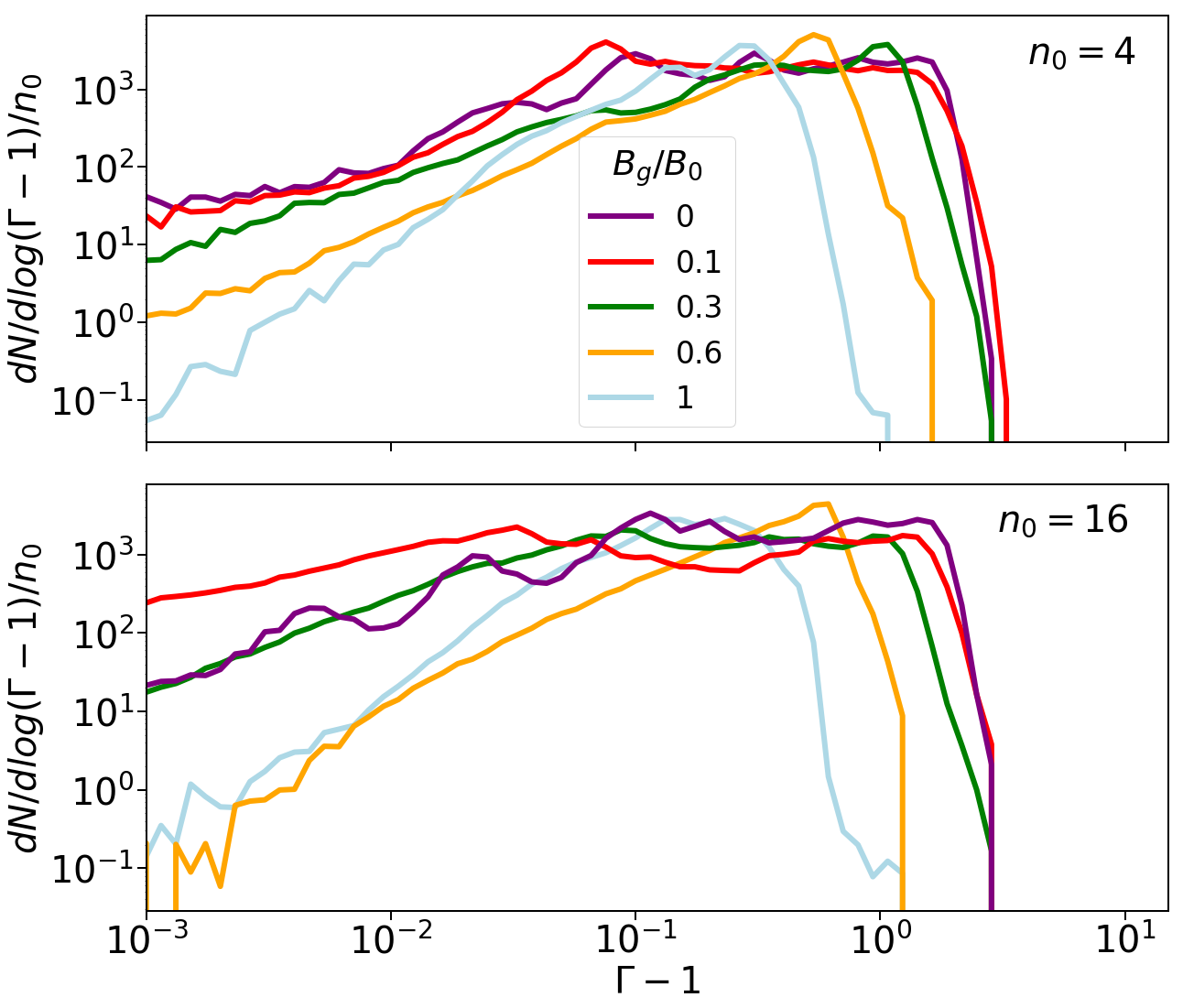}
 \caption{Bulk energy spectra averaged over $2 \leq Tv_{\rm A}/L_{\rm x} \leq 4.2$, for a simulation domain size of $L_{\rm x}/(c/\omega_{\rm p}) = 1680$ and magnetization $\sigma = 10$. The colors represent the guide field strength ($B_{\rm g}/B_0$) and are as follows: purple = 0, red = 0.1, green = 0.3, yellow = 0.6, blue = 1. Top: 4 particles per cell; Bottom: 16 particles per cell. Spectra are normalized by the initial particle density, $n_0$.}
 \label{fig:$n_0$_hist}
\end{figure}

We use $L_{\rm x}/(c/\omega_{\rm p}) = 1680$ and $\sigma = 10$ as our fiducial case for studying the effect of $n_0$ on bulk motions. Fig.~\ref{fig:$n_0$_hist} demonstrates that the spectral features and trends for different guide field cases are similar for $n_0=4$ and $n_0=16$. Comparing the top and bottom panels, we barely notice any difference in the spectra, with the overall shape being consistent for different $n_0$. 

\begin{figure}
 \includegraphics[width=0.5\textwidth]{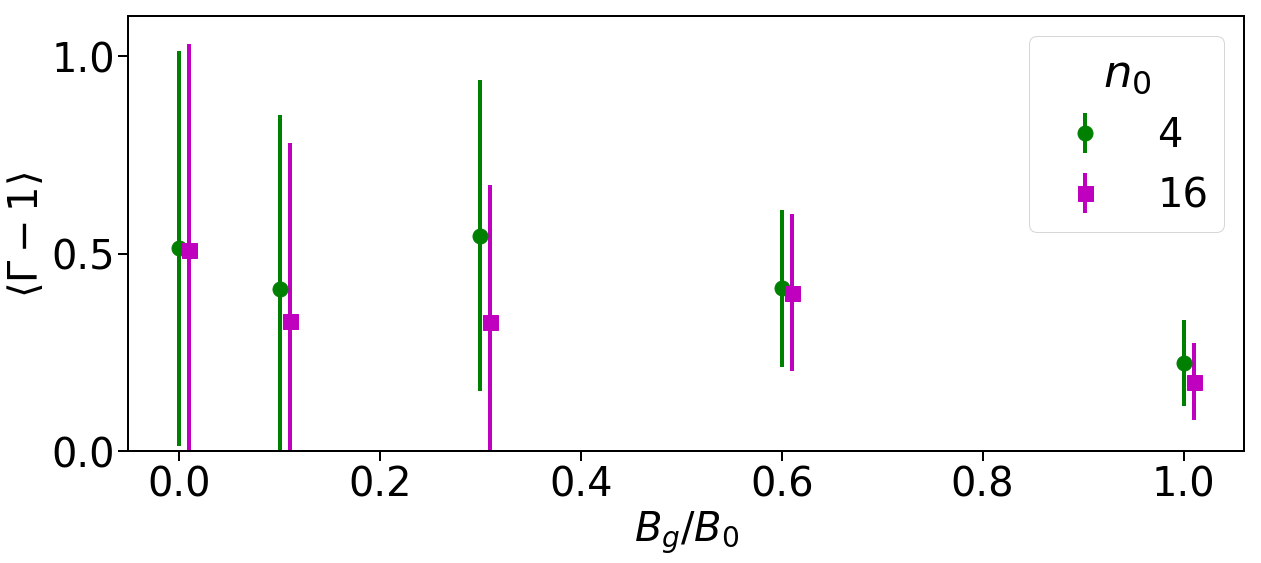}
 \caption{Dependence of the bulk motion energies on different guide fields and different number of particles per cell: green circles and purple squares are for $n_0 = 4$ and $n_0 = 16$, respectively. The error bars indicate the standard deviation of $\langle{\Gamma-1}\rangle$. All means and standard deviations are computed by averaging over $2 \leq Tv_{\rm A}/L_{\rm x} \leq 4.2$.}
 \label{fig:$n_0$}
\end{figure}

Fig.~\ref{fig:$n_0$} shows the effect of different particle densities at initialization on the plasmoid bulk energy. Overall, we notice that both choices of $n_0$ follow the general downward trend we observed in Fig.~\ref{fig:bulkout}. This is further supported by the overlap between data points for $n_0 = 4$ and 16 for strong guide fields. We also notice similar standard deviations for $n_0 = 4$ and 16 across the whole range of guide field strengths, suggesting that there is little dependence on $n_0$. Thus, we conclude that the results shown in the main text for $n_0 = 4$ are robust.

\section{Simulation Parameters} \label{sim_table}

Table \ref{tab:parameters} displays the input parameters of all the simulations presented in the main text as well as some of the output parameters.

\begin{table*}
 \subcaption{Table of numerical and physical parameters}
 \begin{center}
 \begin{tabular}{cccc|cccr}
		\hline
 \hline
		\relax ${B_{\rm g}/B_{0}}^{[1]}$ & ${\sigma}^{[2]}$ & ${L_{\rm x}/(c/\omega_{\rm p})}^{[3]}$ & ${n_0}^{[4]}$ & ${\langle{\Gamma -1}\rangle}^{[5]}$ & ${\Sigma_{\Gamma-1}}^{[6]}$ & {Reconnection Rate,} $v_{\rm in}/v_{\rm A}^{[7]}$\\
		\hline
 \hline 
		0 & 3 & 1680 & 4 & 0.582 & 0.210 & 0.114\\
		0 & 10 & 1680 & 4 & 0.880 & 0.525 & 0.135\\
		0 & 40 & 1680 & 4 & 0.579 & 0.790 & 0.171\\
 \hline
		0.1 & 3 & 1680 & 4 & 0.559 & 0.205 & 0.103\\
 0.1 & 10 & 1680 & 4 & 0.596 & 0.479 & 0.119\\
 0.1 & 40 & 1680 & 4 & 0.391 & 0.588 & 0.135\\
		\hline
 0.3 & 3 & 1680 & 4 & 0.477 & 0.207 & 0.086\\
 0.3 & 10 & 1680 & 4 & 0.597 & 0.426 & 0.091\\
 0.3 & 40 & 1680 & 4 & 0.605 & 0.495 & 0.103\\
 \hline
		0.6 & 3 & 1680 & 4 & 0.366 & 0.156 & 0.056\\
 0.6 & 10 & 1680 & 4 & 0.440 & 0.212 & 0.060\\
 0.6 & 40 & 1680 & 4 & 0.442 & 0.217 & 0.067\\
 \hline
 1 & 3 & 1680 & 4 & 0.177 & 0.115 & 0.033\\
		1 & 10 & 1680 & 4 & 0.198 & 0.123 & 0.038\\
 1 & 40 & 1680 & 4 & 0.208 & 0.120 & 0.042\\
		\hline
 0 & 10 & 1680 & 16 & 0.509 & 0.522 & 0.146\\
		0.1 & 10 & 1680 & 16 & 0.329 & 0.452 & 0.129\\
 0.3 & 10 & 1680 & 16 & 0.326 & 0.346 & 0.097\\
 0.6 & 10 & 1680 & 16 & 0.401 & 0.198 & 0.071\\
 1 & 10 & 1680 & 16 & 0.176 & 0.099 & 0.042\\
 \hline
 0 & 10 & 840 & 4 & 0.523 & 0.506 & 0.151\\
		0.1 & 10 & 840 & 4 & 0.416 & 0.447 & 0.129\\
 0.3 & 10 & 840 & 4 & 0.552 & 0.398 & 0.105\\
 0.6 & 10 & 840 & 4 & 0.417 & 0.200 & 0.071\\
 1 & 10 & 840 & 4 & 0.226 & 0.110 & 0.045\\
		\hline
	\end{tabular}
 \end{center}
 \label{tab:parameters}
 \begin{tablenotes}
 \small
 \item \relax \textit{Note}: All simulations are performed for the same duration of $t_{\rm sim} \sim 4.2\,L_{\rm x}/v_{\rm A}$, with the same spatial resolution of 5 cells per $c/\omega_{\rm p}$. The description of each column is as follows: ${^{[1]}}$ strength of the guide field $B_{\rm g}$ normalized to $B_0$; ${^{[2]}}$ magnetization in the upstream plasma; ${^{[3]}}$ half-length of the computational domain in units of $c/\omega_{\rm p}$; ${^{[4]}}$ initial particle number density in the upstream; ${^{[5]}}$ time- and density-averaged bulk energy in units of rest mass energy; ${^{[6]}}$ standard deviation of bulk energy; ${^{[7]}}$ average reconnection rate during the quasi-steady state (see Fig.~\ref{fig:rec_rate}).
 \end{tablenotes}
\end{table*} 

\section{Simulation Box Size} \label{box}

\begin{figure}
 \includegraphics[width=0.5\textwidth]{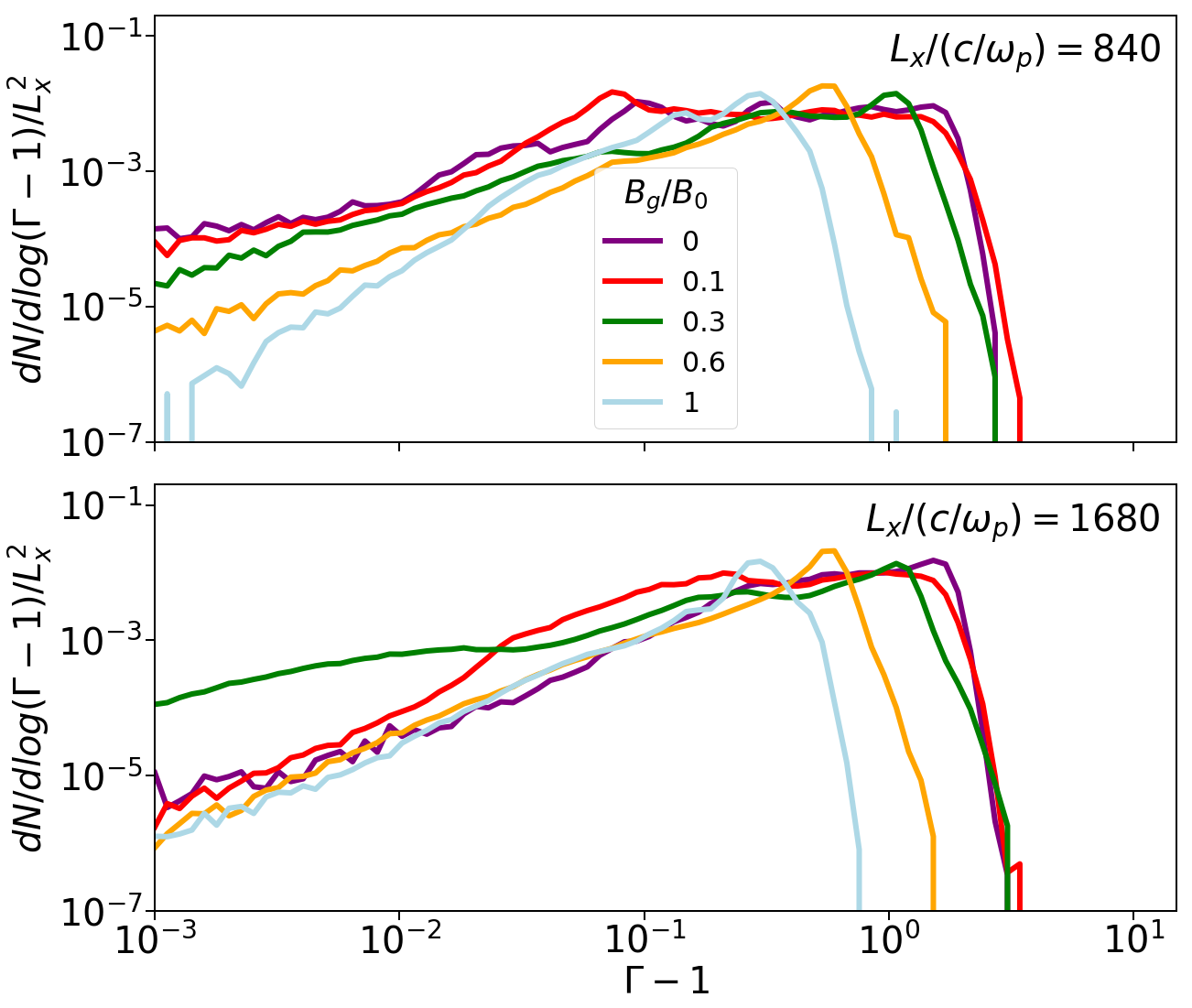}
 \caption{Bulk energy spectra averaged over $2 \leq Tv_{\rm A}/L_{\rm x} \leq 4.2$, for $n_0 = 4$ and $\sigma = 10$. The colors represent the guide field strength ($B_{\rm g}/B_0$) and are as follows: purple = 0, red = 0.1, green = 0.3, yellow = 0.6, blue = 1. Top: simulation domain size of $L_{\rm x}/(c/\omega_{\rm p}) = 840$; Bottom: simulation domain size of $L_{\rm x}/(c/\omega_{\rm p}) = 1680$.}
 \label{fig:box_hist}
\end{figure}

Fig.~\ref{fig:box_hist} shows the bulk energy spectra for different box sizes (at fixed $\sigma = 10$ and $n_0 = 4$): we consider $L_{\rm x}/(c/\omega_{\rm p}) = 840$ and $L_{\rm x}/(c/\omega_{\rm p}) = 1680$ (top and bottom panels, respectively)–––the latter is the reference domain size used in the main text of this paper. We present these data to convey that $L_{\rm x}/(c/\omega_{\rm p}) = 1680$ is sufficiently large for the purposes of our study. Comparing the top and bottom panels, we notice only minor differences in the spectra, while the main trends remain.

\begin{figure}
 \includegraphics[width=0.5\textwidth]{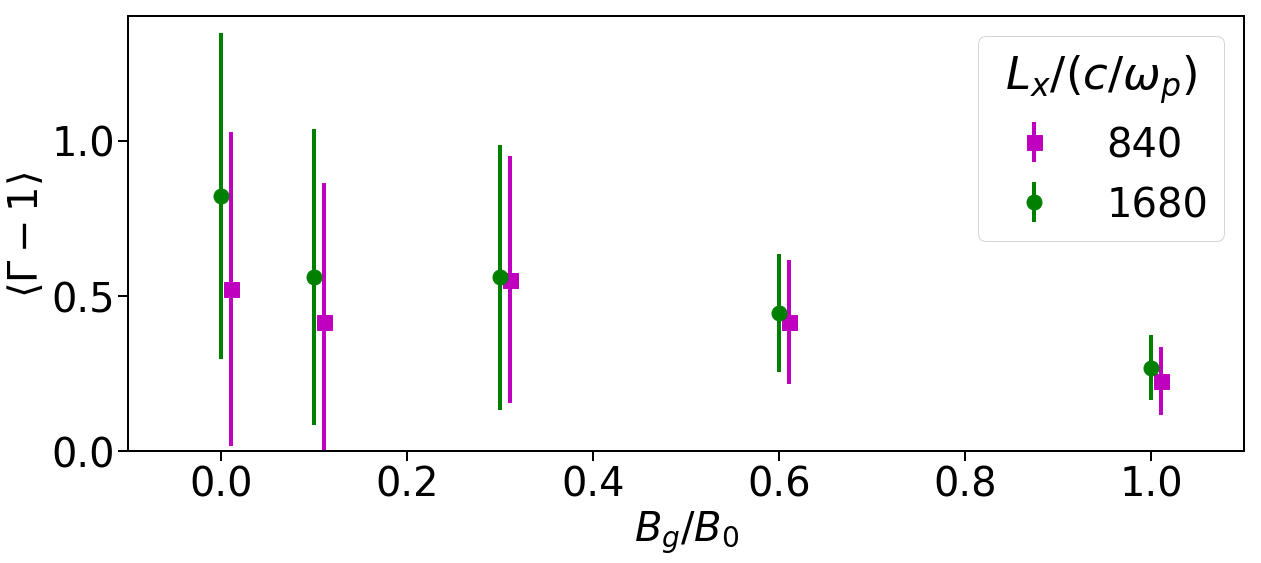}
 \caption{Dependence of the bulk motion energies on different guide fields and different sizes of the simulation box: purple squares and green circles represent $L_{\rm x}/(c/\omega_{\rm p}) = 840$ and $L_{\rm x}/(c/\omega_{\rm p}) = 1680$, respectively. We fix $\sigma=10$. The error bars indicate the standard deviation of $\langle{\Gamma-1}\rangle$. All means and standard deviations are computed by averaging over $2 \leq Tv_{\rm A}/L_{\rm x} \leq 4.2$.} 
 \label{fig:8k}
\end{figure}

Fig.~\ref{fig:8k} shows the effect of varying box size on the average bulk energy and the stochasticity in the plasmoid motions. Overall, we notice that both box sizes follow a similar downward trend in $\langle \Gamma - 1 \rangle$ with increasing guide fields strength. In fact, the values of $\langle \Gamma - 1\rangle$ are almost equal between the two box sizes, for strong guide fields. Finally, we notice consistency in the level of stochasticity in bulk motions from the size of the error bars. This suggests that our results are converged with respect to box size.

\section{Simulation Boundaries} \label{boundary}

\begin{figure*}
 \includegraphics[width=\textwidth]{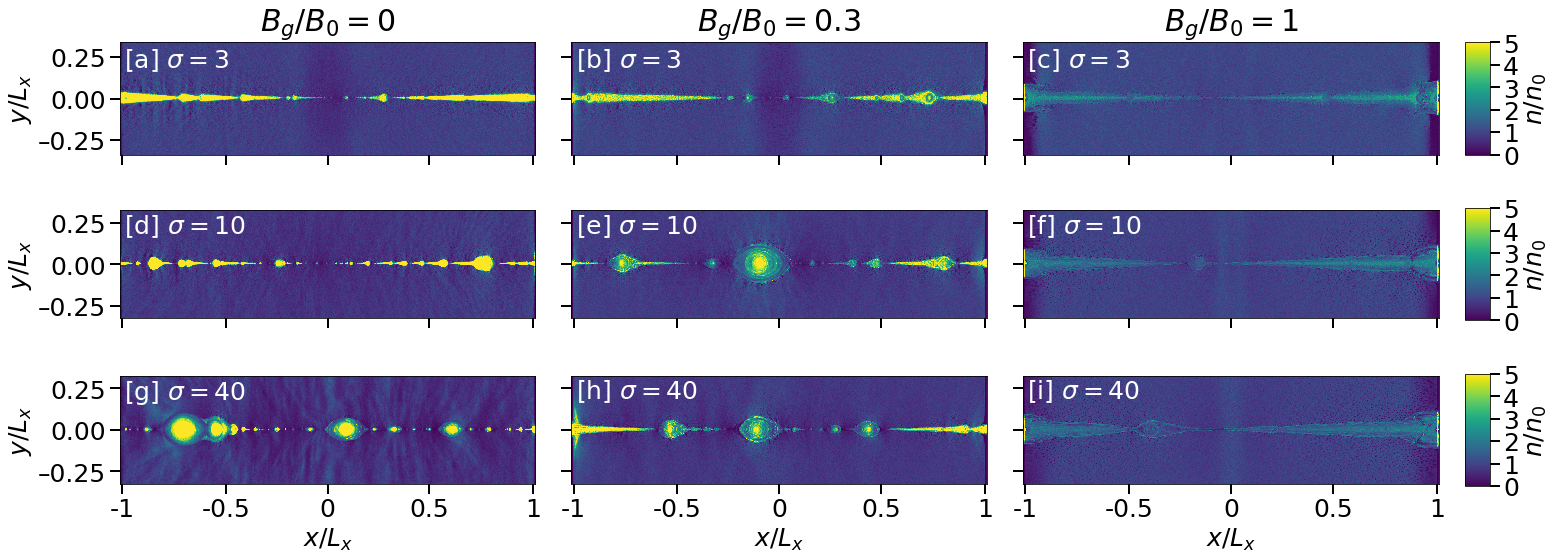}
 \caption{2D snapshots of the reconnection layer at time $Tv_{\rm A}/L_{\rm x} \simeq 4$ for magnetizations increasing from top to bottom ($\sigma = 3, \sigma = 10, \sigma = 40$) and guide fields increasing from left to right ($B_{\rm g}/B_0 = 0, B_{\rm g}/B_0 = 0.3, B_{\rm g}/B_0 = 1$) without removing the cells adjacent to the outflow boundary walls (as done, instead, in Fig.~\ref{fig:2d_img}). The figures display the normalized particle number density, $n/n_0$.}
 \label{fig:og2d}
\end{figure*}

In simulations with a strong guide field, our outflow boundary conditions are not able to optimally advect the compressed guide field in the reconnected plasma out of the box. This leads to a gradual increase in the guide field strength near the boundaries, which inhibits a perfectly smooth exhaust of the outflowing plasma. As a result, we notice clumping of plasma along the $x$-edges of the simulation box (see panels [c, f, i] of Fig.~\ref{fig:og2d}). To overcome this spurious effect, all the analyses in the main paper excluded the simulation cells in the vicinity of the $x$-boundaries (more precisely, within a distance of $0.08\,L_{\rm x}$ from each boundary). 

\section{Assessment of the Quasi-Steady State} \label{steady_state}
Throughout this paper, many of the results are obtained by taking the time average over the quasi-steady state, defined as $2 \leq Tv_{\rm A}/L_{\rm x} \leq 4.2$. This spans the range from the time when the two reconnection fronts have exited the box, until the end of our simulations.

\begin{figure*}
 \includegraphics[width=\textwidth]{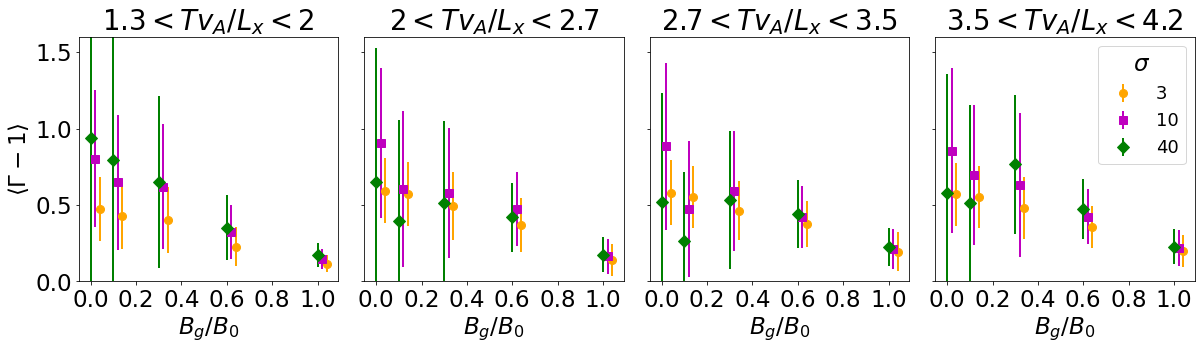}
 \caption{Time-averaged plots of mean and standard deviation of $\Gamma - 1$ derived using bulk spectra as in Fig.~\ref{fig:bulk_spect}, but focusing on different time ranges. Left: $1.3 \leq Tv_{\rm A}/L_{\rm x} \leq 2$; Left-middle: $2 \leq Tv_{\rm A}/L_{\rm x} \leq 2.7$; Right-middle: $2.7 \leq Tv_{\rm A}/L_{\rm x} \leq 3.5$; Right: $3.5 \leq Tv_{\rm A}/L_{\rm x} \leq 4.2$. Yellow circles, purple squares, and green diamonds refer to $\sigma = 3$, $\sigma = 10$, and $\sigma = 40$, respectively.}
 \label{fig:time_bulkout}
\end{figure*}

In Fig.~\ref{fig:time_bulkout}, we sub-divide the range $1.3 \leq Tv_{\rm A}/L_{\rm x} \leq 4.2$ into four time intervals ––– $1.3 \leq Tv_{\rm A}/L_{\rm x} \leq 2$, $2 \leq Tv_{\rm A}/L_{\rm x} \leq 2.7$, $2.7 \leq Tv_{\rm A}/L_{\rm x} \leq 3.5$, and $3.5 \leq Tv_{\rm A}/L_{\rm x} \leq 4.2$ ––– and compute both the average, $\langle{\Gamma -1}\rangle$, and the standard deviation, $\Sigma_{\Gamma-1}$, for each time range. We choose to include the first panel, which refers to $1.3 \leq Tv_{\rm A}/L_{\rm x} \leq 2$, to check whether our conclusions would be different, if we were to consider a time interval before the establishment of the steady state (at $Tv_{\rm A}/L_{\rm x} \leq 2 $, the two outflowing reconnection fronts have yet to exit the domain). 


Overall, we notice trends consistent with those in Fig.~\ref{fig:bulkout} as well as Fig.~\ref{fig:8k} and Fig.~\ref{fig:$n_0$}, with a general decrease in mean bulk energy as the guide field increases. This trend is observed in all the time ranges (i.e., all panels). Thus, Fig.~\ref{fig:time_bulkout} convincingly demonstrates that our results are robust, and that the spurious accumulation of plasma at the $x$-boundaries (which worsens with time, as discussed in the main text) does not impact the main trends in the bulk motion properties (the results from the last three panels are nearly identical).

\section{Momentum Space Plots} \label{bulk_cont}

\begin{figure*}
 \includegraphics[width=\textwidth]{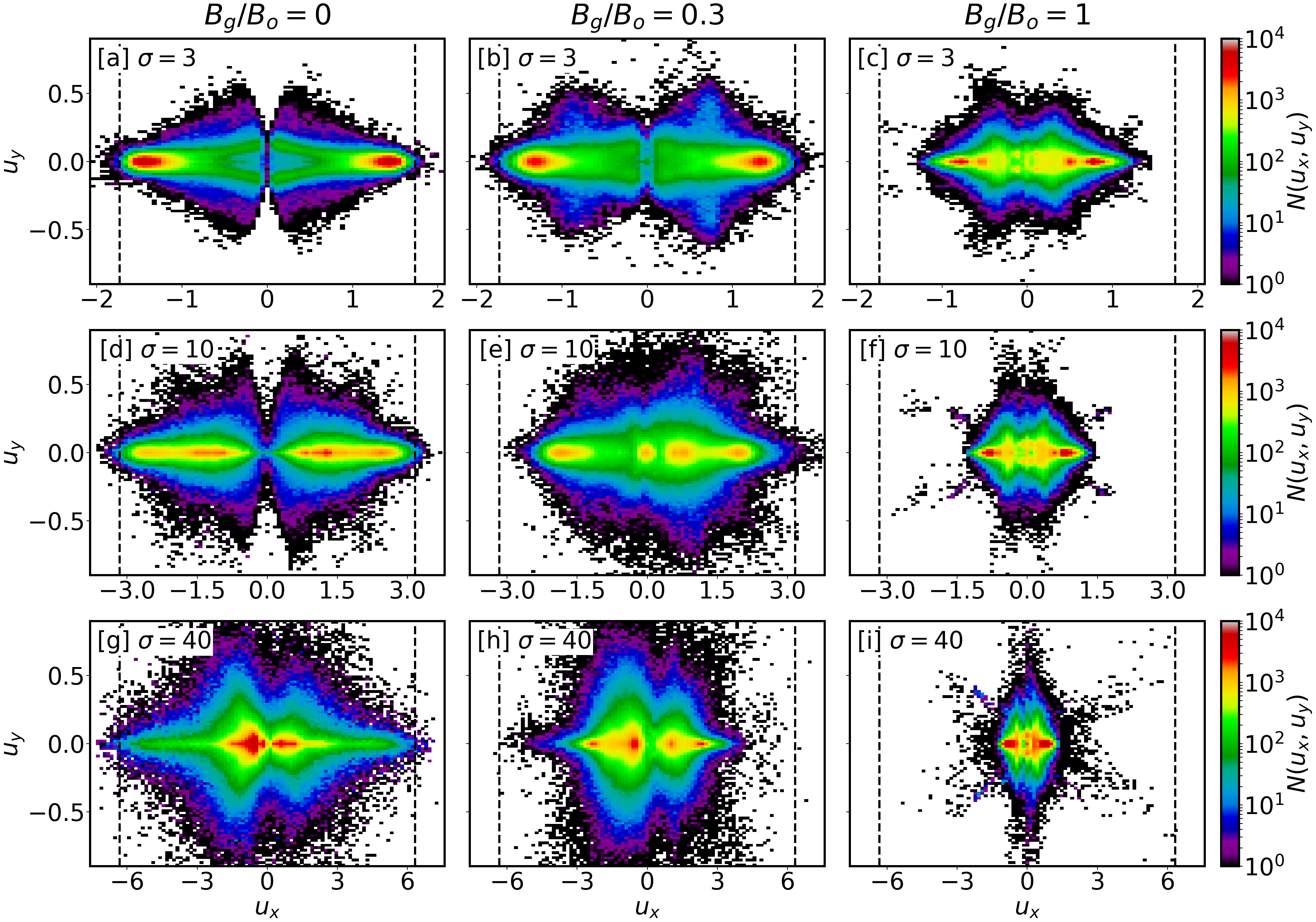}
 \caption{Bulk motions of the reconnected plasma, viewed in the $u_{\rm x}-u_{\rm y}$ phase space. The color represents the particle number density in phase space. The figures are arranged as follows: magnetization increases from top to bottom ($\sigma = 3, \sigma = 10, \sigma = 40$) and guide field increases from left to right ($B_{\rm g}/B_0 = 0, B_{\rm g}/B_0 = 0.3, B_{\rm g}/B_0 = 1$). The dotted vertical lines show the Alfv\'{e}nic limit, $\sqrt{\sigma}$. All phase space plots are time-averaged over $2 \leq Tv_{\rm A}/L_{\rm x} \leq 4.2$, when the reconnection layer is in a quasi-steady state.}
 \label{fig:uxuy}
\end{figure*}

In Fig.~\ref{fig:uxuy}, we use bulk 4-velocities both parallel and orthogonal to the reconnection layer, respectively $u_{\rm x} = \Gamma\beta_{\rm x}$ (along the outflow direction) and $u_{\rm y} = \Gamma\beta_{\rm y}$ (along the inflow direction).

We confirm the two main results of Fig.~\ref{fig:yux}: (1) the outflow motions (i.e., in $u_{\rm x}$) are slower for stronger guide fields; (2) a smaller fraction of the reconnected plasma reaches the Alfv\'{e}nic limit $|u_{\rm x}|\sim \sqrt{\sigma}$ at higher magnetizations. 
Bulk motions of the reconnected plasma along $y$ are expected as a result of secondary current sheets formed perpendicular to the primary current sheet, at the interface of merging plasmoids. Generally, we find that bulk speeds along $y$ are much smaller than along $x$. For a given magnetization, we find the average $\lvert u_{\rm y}/u_{\rm x}\rvert$ increases with stronger guide fields. This effect is less so due to an increase in $u_{\rm y}$, and more so due to the decrease in $u_{\rm x}$ with increasing guide field strength (e.g., see Fig.~\ref{fig:yux}). 

These results are consistent with previous papers of this series \citet{SB20, SSB21}, in the regime of negligible cooling. \citet{SB20, SSB21}, on the other hand, found faster bulk motions along $y$ (comparable to $u_{\rm x}$) in strongly cooled simulations. This is likely related to the effective magnetization of plasmoids (which, in the case of merging plasmoids, serve as the upstream regions for the current sheet at the merger interface). In the uncooled case, plasmoids have comparable magnetic and thermal energies, so their effective magnetization is around unity (here, we normalize the magnetic field enthalpy density to the overall plasma enthalpy density, including thermal contributions). In a strongly cooled case, instead, the effective magnetization is comparable to the value of the primary layer, which leads to faster $y$-directed bulk motions. 


\bsp	
\label{lastpage}
\end{document}